\documentclass[preprint,12pt,3p]{elsarticle}
\usepackage{amssymb}
\usepackage{lineno}
\usepackage{graphicx}
\usepackage[caption=false]{subfig}
\usepackage{hyperref} %
\usepackage[usenames,dvipsnames]{xcolor}
\usepackage[caption=false]{subfig}
\hypersetup{colorlinks=true, urlcolor=blue, linkcolor=blue, citecolor=red} 

\begin{document}

\title{The COSINE-100 Liquid Scintillator Veto System}

\author[a,q]{G.~Adhikari}
\address[a]{Department of Physics, University of California San Diego, La Jolla, CA 92093, USA}

\author[b]{E.~Barbosa~de~Souza}
\address[b]{Department of Physics and Wright Laboratory, Yale University, New Haven, CT 06520, USA}

\author[c]{N.~Carlin}
\address[c]{Physics Institute, University of S\~{a}o Paulo, 05508-090, S\~{a}o Paulo, Brazil}

\author[d]{J.~J.~Choi}
\address[d]{Department of Physics and Astronomy, Seoul National University, Seoul 08826, Republic of Korea}

\author[d]{S.~Choi}

\author[e]{M.~Djamal}
\address[e]{Department of Physics, Bandung Institute of Technology, Bandung 40132, Indonesia}

\author[f]{A.~C.~Ezeribe}
\address[f]{Department of Physics and Astronomy, University of Sheffield, Sheffield S3 7RH, United Kingdom}

\author[c]{L.~E.~Fran{\c c}a}

\author[g]{C.~Ha}
\address[g]{Department of Physics, Chung-Ang University, Seoul 06973, Republic of Korea}

\author[h,i,j]{I.~S.~Hahn}
\address[h]{Department of Science Education, Ewha Womans University, Seoul 03760, Republic of Korea} 
\address[i]{Center for Exotic Nuclear Studies, Institute for Basic Science (IBS), Daejeon 34126, Republic of Korea}
\address[j]{IBS School, University of Science and Technology (UST), Daejeon 34113, Republic of Korea}

\author[k]{E.~J.~Jeon}
\address[k]{Center for Underground Physics, Institute for Basic Science (IBS), Daejeon 34126, Republic of Korea}

\author[b]{J.~H.~Jo}

\author[d]{H.~W.~Joo}

\author[k]{W.~G.~Kang}

\author[l]{M.~Kauer}
\address[l]{Department of Physics and Wisconsin IceCube Particle Astrophysics Center, University of Wisconsin-Madison, Madison, WI 53706, USA}

\author[k]{H.~Kim}

\author[m]{H.~J.~Kim}
\address[m]{Department of Physics, Kyungpook National University, Daegu 41566, Republic of Korea}

\author[k]{K.~W.~Kim}

\author[d]{S.~K.~Kim}

\author[j,k]{W.~K.~Kim}

\author[j,k,n]{Y.~D.~Kim}
\address[n]{Department of Physics, Sejong University, Seoul 05006, Republic of Korea}

\author[j,k,o]{Y.~H.~Kim}
\address[o]{Korea Research Institute of Standards and Science, Daejeon 34113, Republic of Korea}

\author[k]{Y.~J.~Ko}

\author[k]{E.~K.~Lee}

\author[j,k]{H.~Lee}

\author[j,k]{H.~S.~Lee}

\author[k]{H.~Y.~Lee}

\author[k]{I.~S.~Lee}

\author[k]{J.~Lee}

\author[m]{J.~Y.~Lee}

\author[j,k]{M.~H.~Lee}

\author[j,k]{S.~H.~Lee}

\author[d]{S.~M.~Lee}

\author[k]{D.~S.~Leonard}

\author[c]{B.~B.~Manzato}

\author[b]{R.~H.~Maruyama}

\author[f]{R.~J.~Neal}

\author[k]{S.~L.~Olsen}

\author[j,k]{B.~J.~Park}

\author[q]{H.~K.~Park}
\address[q]{Department of Accelerator Science, Korea University, Sejong 30019, Republic of Korea}

\author[o]{H.~S.~Park}

\author[k]{K.~S.~Park}

\author[c]{R.~L.~C.~Pitta}

\author[k]{H.~Prihtiadi}

\author[k]{S.~J.~Ra}

\author[r]{C.~Rott}
\address[r]{Department of Physics, Sungkyunkwan University, Suwon 16419, Republic of Korea}

\author[k]{K.~A.~Shin}

\author[f]{A.~Scarff}

\author[f]{N.~J.~C.~Spooner}

\author[b]{W.~G.~Thompson}

\author[a]{L.~Yang}

\author[r]{G.~H.~Yu}

%\date{\today}

\begin{abstract}
This paper describes the liquid scintillator veto system for the COSINE-100 dark matter experiment and its performance. The COSINE-100 detector consists of eight NaI(Tl) crystals immersed in 2200~L of linear alkylbenzene-based liquid scintillator. The liquid scintillator tags between 65 and 75\% of the internal $^{40}$K background in the 2--6~keV energy region. We also describe the background model for the liquid scintillator, which is primarily used to assess its energy calibration and threshold.
\end{abstract}

\maketitle
%\linenumbers

\section{Introduction}\label{sec:intro}
Overwhelming cosmological and astronomical observations provide strong evidence that the most of the matter in the universe consists of non-relativistic dark matter~\cite{Ade:2013zuv, Bertone:2004pz}. The weakly interacting massive particle (WIMP) is one of the theoretically favored candidates to explain this dark matter~\cite{Steigman:1984ac}. 

COSINE-100~\cite{COSINE_detector} is a direct dark matter detection experiment to test the DAMA experiment's observation of an annual modulation signal~\cite{Bernabei:2005hj,Bernabei:2013xsa,Bernabei:2018} using the same target material as DAMA. The COSINE-100 detector consists of eight NaI(Tl) crystals with a total mass of 106~kg, used as active target. The crystals are immersed in 2200~L of liquid scintillator (LS) surrounded by 37 plastic scintillator panels.

It is critical to reduce the background as much as possible for dark matter searches because of the low cross-section of the WIMP-nucleon interaction. The dominant background contributions come from radioactivity in detector materials surrounding the NaI(Tl) detectors and backgrounds internal to the NaI(Tl) detectors themselves.

The COSINE-100 experiment utilizes the LS systems as active vetos to reduce the background level in the NaI(Tl) detectors by tagging neutrons and $\gamma$ events from external sources. In addition, the LS veto can also identify radioactive background events internal to the crystal detectors by tagging escaping $\gamma$-rays. The data used in this report were acquired between 20 October 2016 and 19 December 2016, with a total exposure of 59.5 live days. During this period, no substantial environmental anomalies or detector instabilities, such as temperature, humidity, or detector voltage/current fluctuations were observed.

\section{Liquid Scintillator Veto System Configuration}\label{sec:setup}

\begin{figure*}[!h]
\label{fig:detectors}%
\centering
    \subfloat[Detector Configuration]{%
        \centering
        \includegraphics[width=0.5\columnwidth]{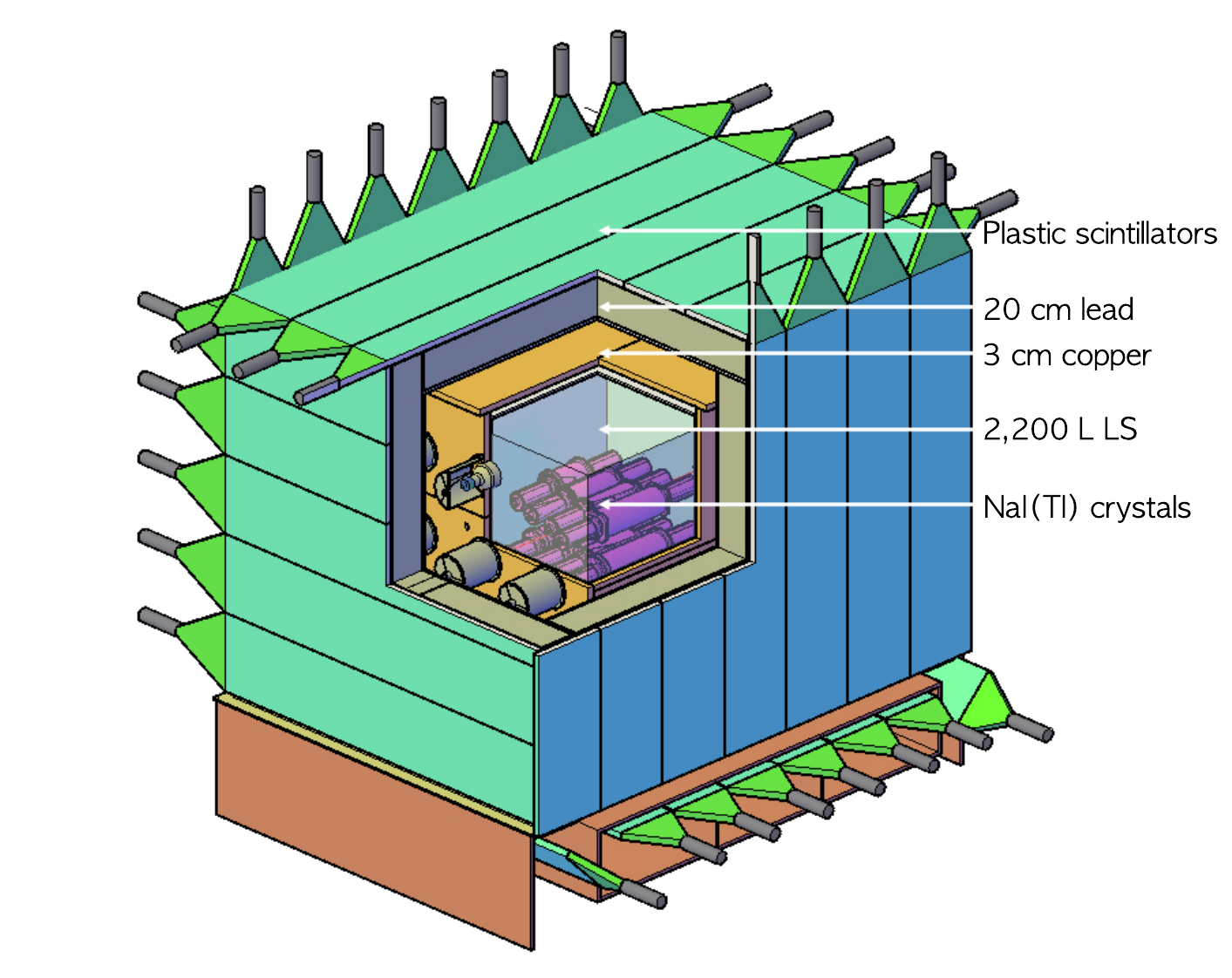}
        \label{fig:detector}%
    }
    \subfloat[Detector Side view]{%
        %\centering
        \includegraphics[width=0.5\columnwidth]{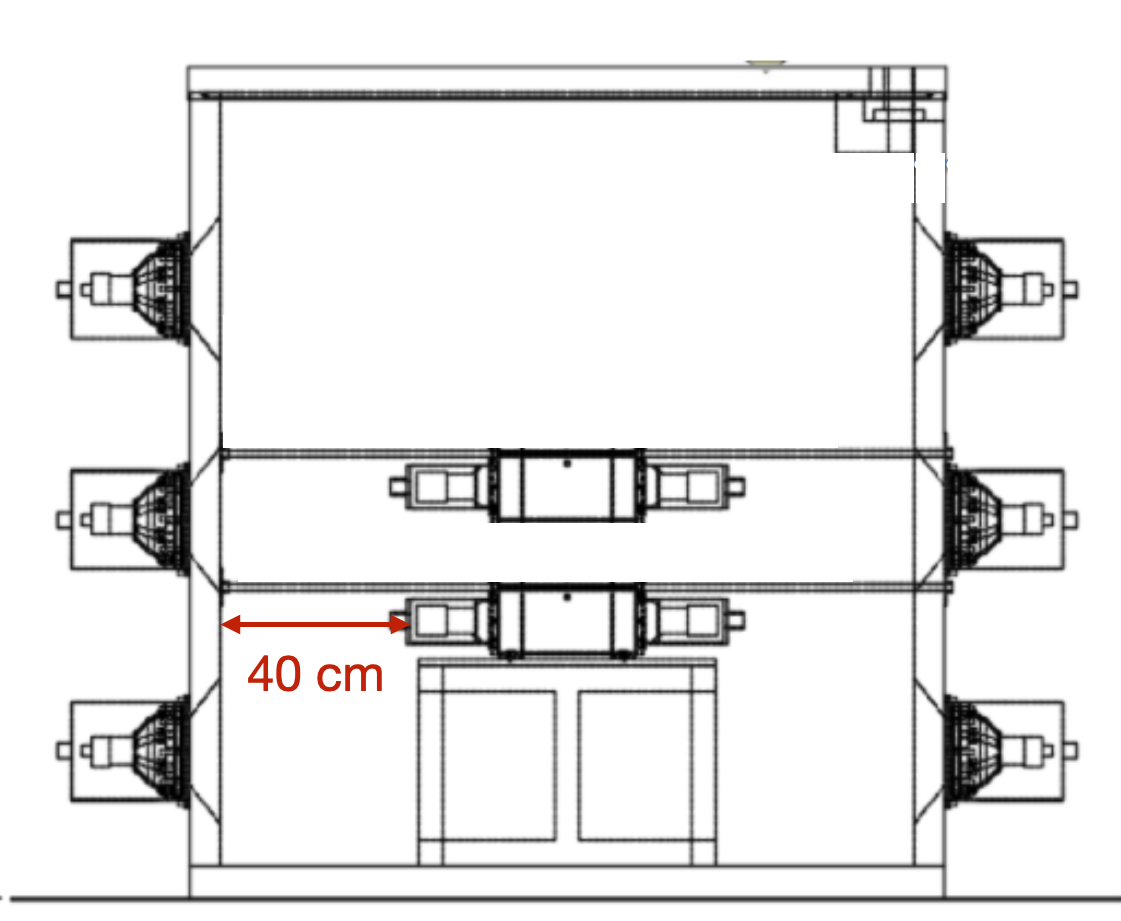}
        \label{fig:sideview}%
    }
    
\caption{ (a) A schematic of the COSINE-100 detector system. 2,200~L of liquid scintillator is filled within an acrylic container inside of a copper shield. The top 9~cm of the acrylic container is left unfilled and nitrogen gas is circulated in this volume. (b) A side view of the detector system. The minimum distance between the PMTs of the crystals and the inner wall of the acrylic container is approximately 40~cm.}
\end{figure*}

The liquid scintillator in the COSINE-100 detector is held within in an acrylic container. The inner walls of the acrylic container and the outer surfaces of the crystal assemblies are wrapped with Vikuiti-ESR specular reflective films to increase the light collection efficiency. The photons produced in the LS are detected by 18 5-inch R877 photomultiplier tubes (PMTs) from Hamamatsu, installed at two opposite sides of the container. The minimum distance between the PMTs attached to the crystals and the acrylic container inner wall is approximately 40~cm. Figures~\ref{fig:detector} and~\ref{fig:sideview} show a schematic representation of the COSINE-100 detector and shielding structure with the liquid scintillator, and a side view of the detector system, respectively. A more detailed description of the detector can be found in Ref.~\cite{COSINE_detector}.

The top 9~cm of the acrylic container housing the LS are unfilled as a safety margin in the case of a temperature increase which may induce an expansion of the LS volume. Gas boil-off from liquid nitrogen is injected into this volume at a rate of 3~liters per minute to prevent the liquid scintillator contacting oxygen or water vapor, in order to maintain a high light yield. The relative humidity in this region is kept at $<$2.0\% and the high heat capacity of the LS helps keep the temperature of the liquid stable at 24.20$\pm$0.05~$^{\circ}$C.

Signals from the PMTs are collected with two charge-sensitive analog to digital converter (ADC) modules with a sampling rate of 64~MHz~\cite{COSINE_daq}. Each ADC module has 32 channels and its peak-to-peak voltage is 2~V with 12-bit resolution. The signals from the 18 PMTs of the LS veto detector are amplified by a factor of 30. The LS veto data is passively triggered, the ADC modules only take data when the NaI(Tl) crystals generate a trigger. Ref.~\cite{COSINE_daq} describes the details of the trigger algorithm used for the LS veto system.

\section{Production of Liquid Scintillator}\label{sec:production}

COSINE-100 uses a linear alkylbenzene (LAB)-based liquid scintillator~\cite{LABLS1, LABLS2}, which contains 3~g/L of 2,5-diphenyloxazole (PPO) and 30~mg/L of 1,4-bis (2- methylstyryl) benzene (bis-MSB) as a primary and secondary wavelength shifter~\cite{LABLS3}. The emission spectrum of the LAB has a maximum at 340~nm; the wavelength shifter mixed with the solvent match the wavelength of the scintillation light with the maximum efficiency of the PMTs used in the COSINE-100 experiment. The wavelength shifter boosts the optical transparency of the solvent as well. In order to extract any insoluble impurities, the LAB is filtered by Meissner filters that have a 0.1 $\mu$m pore size. After mixing the PPO and bis-MSB, the LS is purified using a water extraction method~\cite{LSWater} to avoid contamination from natural radioisotopes.

Despite the water extraction process for purification, if the LS still has a sufficiently high radioisotope level, such as  $^{238}$U or $^{232}$Th, this will contribute to the background measured in the NaI(Tl) crystals. One needs to carefully measure the background level of the LS so that the background observed in the NaI(Tl) crystals can be accurately characterized. For this purpose, a prototype LS detector ($\sim$70~mL in volume) read out by 3-inch PMTs was installed inside the KIMS shielding facility~\cite{Kim:2012rza} at the Yangyang underground laboratory. Alpha particle signals are separated from gamma-induced signals by the pulse discrimination method~\cite{COSINE_neutron}. Possible external neutron background events which cannot be discriminated from alphas with the available detector resolution is negligible because of the polyethylene shielding surrounding the prototype detector. A time-coincidence analysis is performed between sequential events to estimate the upper limit of the intrinsic $^{238}$U and $^{232}$Th chain activity~\cite{KIMS_NaI}. The beta-alpha coincidence between $^{214}$Bi to $^{214}$Po is used to estimate the $^{238}$U activity and was found to be 0.087 mBq/kg. The alpha-alpha coincidence between $^{222}$Rn and $^{218}$Po is used to estimate the $^{232}$Th activity and was found to be 0.016 mBq/kg. We determine that the background contributions from the LS veto to the NaI(Tl) crystal energy spectra are negligible considering upper limits from the prototype detector based on the simulation studies of the full COSINE-100 detector~\cite{LABLS2}. 

\section{Calibration}\label{sec:calibration}

\begin{figure}[!h]
	\centering
	\includegraphics[width=0.7\columnwidth]{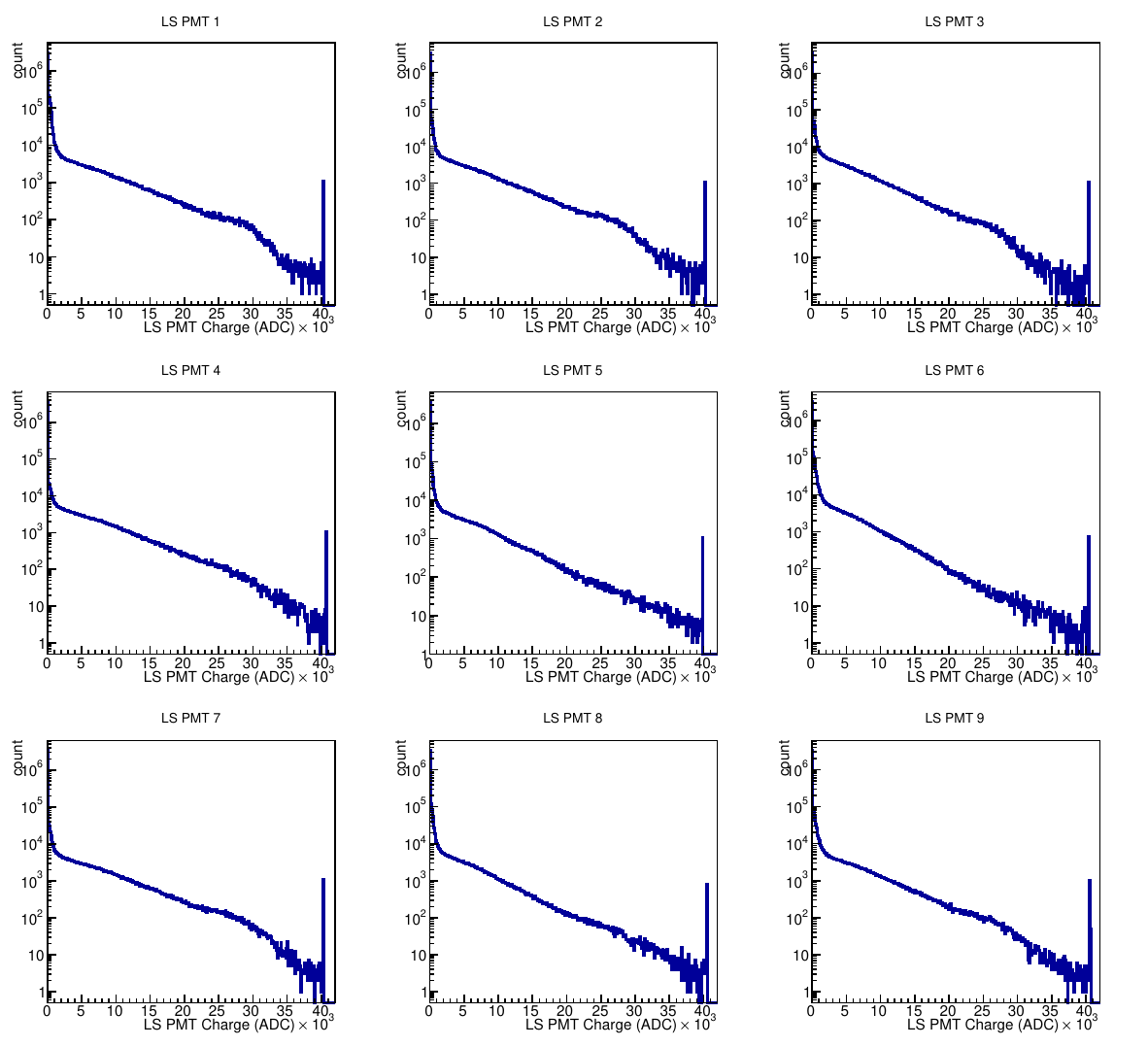}
	\caption{Spectra of nine 5-inch PMTs in one side of the COSINE-100 liquid scintillator veto after gain correction, with visible saturation point. The total integrated charge is calculated by summing the charge of all 18 PMTs.} %Overflow bins at around 40,000~ADC counts due to saturation of the PMTs.} 
	\label{fig:lsCharge_pmt}
\end{figure}

\begin{figure}[!h]
	\centering
	\includegraphics[width=0.5\columnwidth]{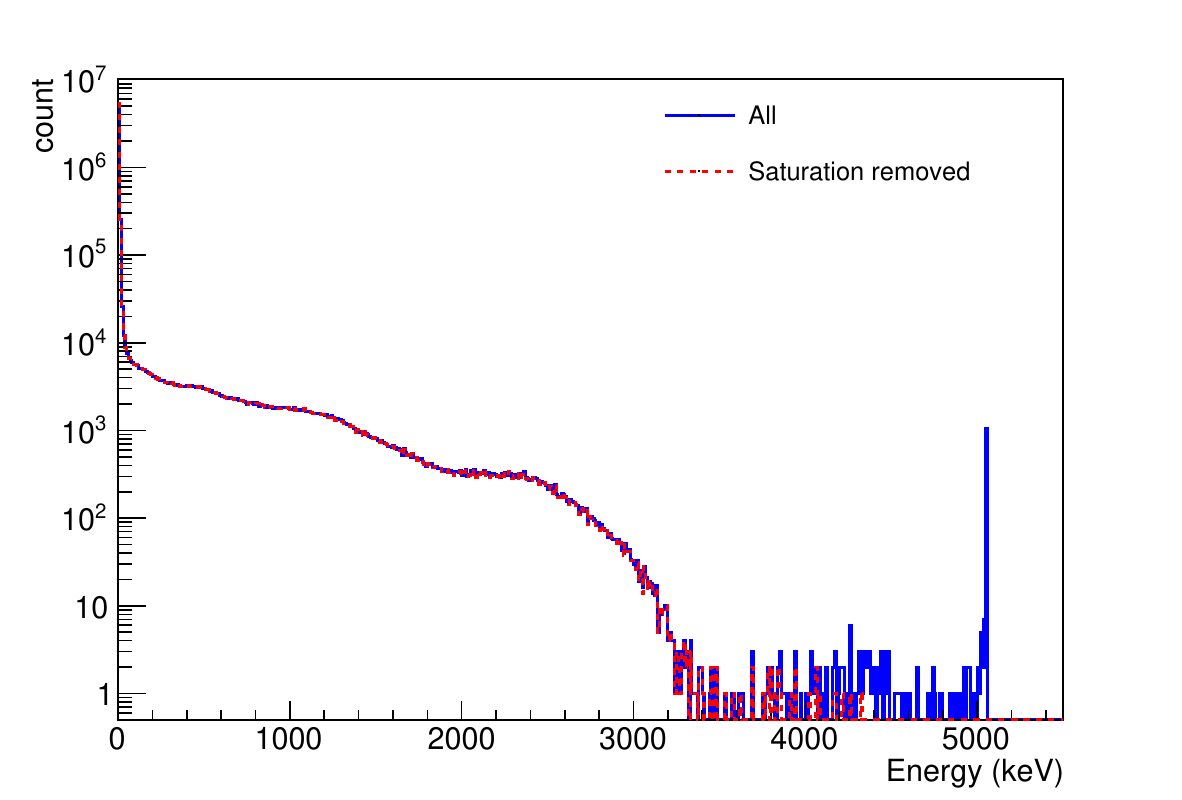}
	\caption{Liquid scintillator veto spectra after calibration. The total liquid scintillator energy spectrum is the sum of the integrated charges from all 18 PMTs. The blue histogram shows the energy spectrum from all events, whereas the red spectrum has had events that saturate at least a single PMT removed. Removing saturated events does not affect the spectral shape below $\sim3.2$ MeV.}
	\label{fig:lsChargeSpectrum}
\end{figure}

The calibration of the 18 PMTs in the LS veto starts with matching the gain of these PMTs, as shown in Figure~\ref{fig:lsCharge_pmt}. The gain was matched such that each PMT saturates at around 40,000~ADC.
The integrated charges of the 18 PMTs are then summed to produce the uncalibrated energy spectrum of the entire LS veto. Events that saturate at least a single PMT are removed to avoid events with misrepresented energies, as seen in Figure~\ref{fig:lsChargeSpectrum}. Note that if an event saturates one PMT, it does not mean that all of the other will also be saturated. As a result of this, we observe that several events bellow the saturation point of the total charge spectrum in Figure~\ref{fig:lsChargeSpectrum} are also removed. 
The spectrum does not present characteristic features, but it is possible to notice two main shoulders at higher energies. These are due to $^{40}$K (1.46~MeV) and $^{208}$Tl (2.6~MeV) Compton scattering. %More about those features will be discussed in the following sections.

\begin{figure}[!h]
	\centering
	\includegraphics[width=0.5\columnwidth]{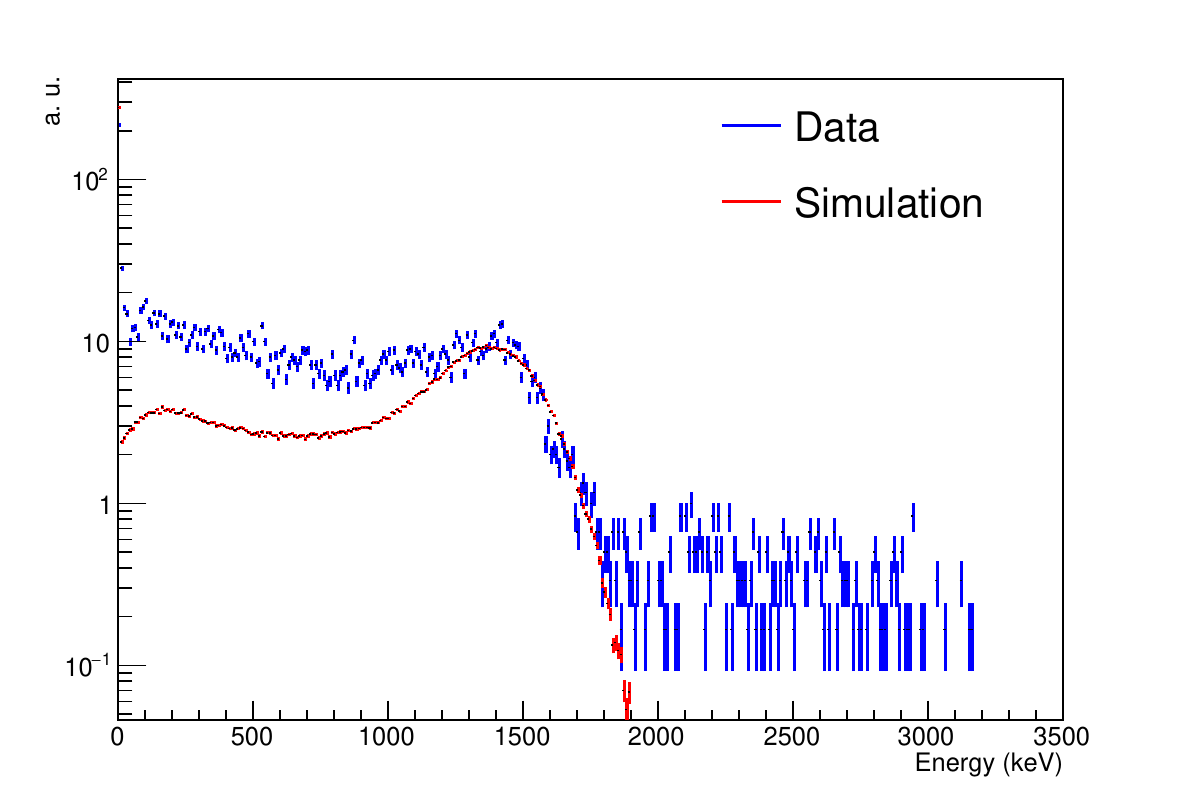}
	\caption{Liquid scintillator veto tagging spectra for crystal events with energies between 2-4 keV. The comparison between data (blue) and simulation (red) is shown at an arbitrary rate scale. The Compton shoulder from 1,460 keV $\gamma$-rays originating from $^{40}$K decays in the crystals is clearly visible in both the data and simulated spectra. The data spectrum has higher rates at lower energies due to other background components that can act at those energies, which are not considered in this simulated $^{40}$K spectrum. At energies above $\sim1.9$ MeV, events cannot be attributed to $^{40}$K only.}%, so they must correspond to decays of other components or wrong coincidence.}
	\label{fig:ls_K40DataMC}
\end{figure}

The LS spectrum does not exhibit mono-energetic features which could be used for energy calibration or estimation of the energy resolution. Therefore, in order to calculate the LS veto calibration factor we compare the data to a simulated spectrum, applying different resolution functions. In the simulated spectrum we consider the energy deposited in the LS by all known detector contaminants. We also compare crystal-LS coincidence spectra, selecting specific classes of events that are tagged by both the crystal detectors and LS. %That is done by comparing the total backgrounds measured for detectors' materials and crystal-LS coincidence spectra to, selecting a specific group of events that could be tagged by both crystal and liquid scintillator. 
For example, we select $^{40}$K events which exhibit coincident 3.2~keV energy depositions from Auger electrons in the NaI(Tl) detectors and higher energy depositions (up to 1,460~keV) in the LS. The shoulder around 1,460~keV is then matched between the simulation and data to determine the LS energy resolution ($\sim10\%$) and calibration. The comparison between this coincidence data and the $^{40}$K simulation is shown in Figure~\ref{fig:ls_K40DataMC}.

\section{Background Fit}\label{sec:background}
Once the calibration and resolution have been determined, we model the liquid scintillator background by fitting the full energy range with a simulated energy spectrum. The Geant4~\cite{geant,geant_2} simulated spectrum consists of components from the NaI(Tl) detectors, PMTs, the LS itself, and other external components, such as the detector shielding structure. Each component is simulated separately. For the dataset used in this analysis the LS veto operated in a passive trigger mode, only recording an event when triggered by an energy deposition in an NaI(Tl) detector. Thus, a cut must be applied to the simulated events to select identify only events that deposit energy in at least one crystal.

The fit is performed via the method of maximum likelihood, taking into consideration the simulated spectra from both internal and external backgrounds~\cite{COSINE_background}, as well as the statistical uncertainties of the data. Only the energy region between 500-3000 keV is fitted, as the resolution was estimated exclusively for these higher energies. 

\begin{figure}[!h]
	\centering
	\includegraphics[width=0.7\columnwidth]{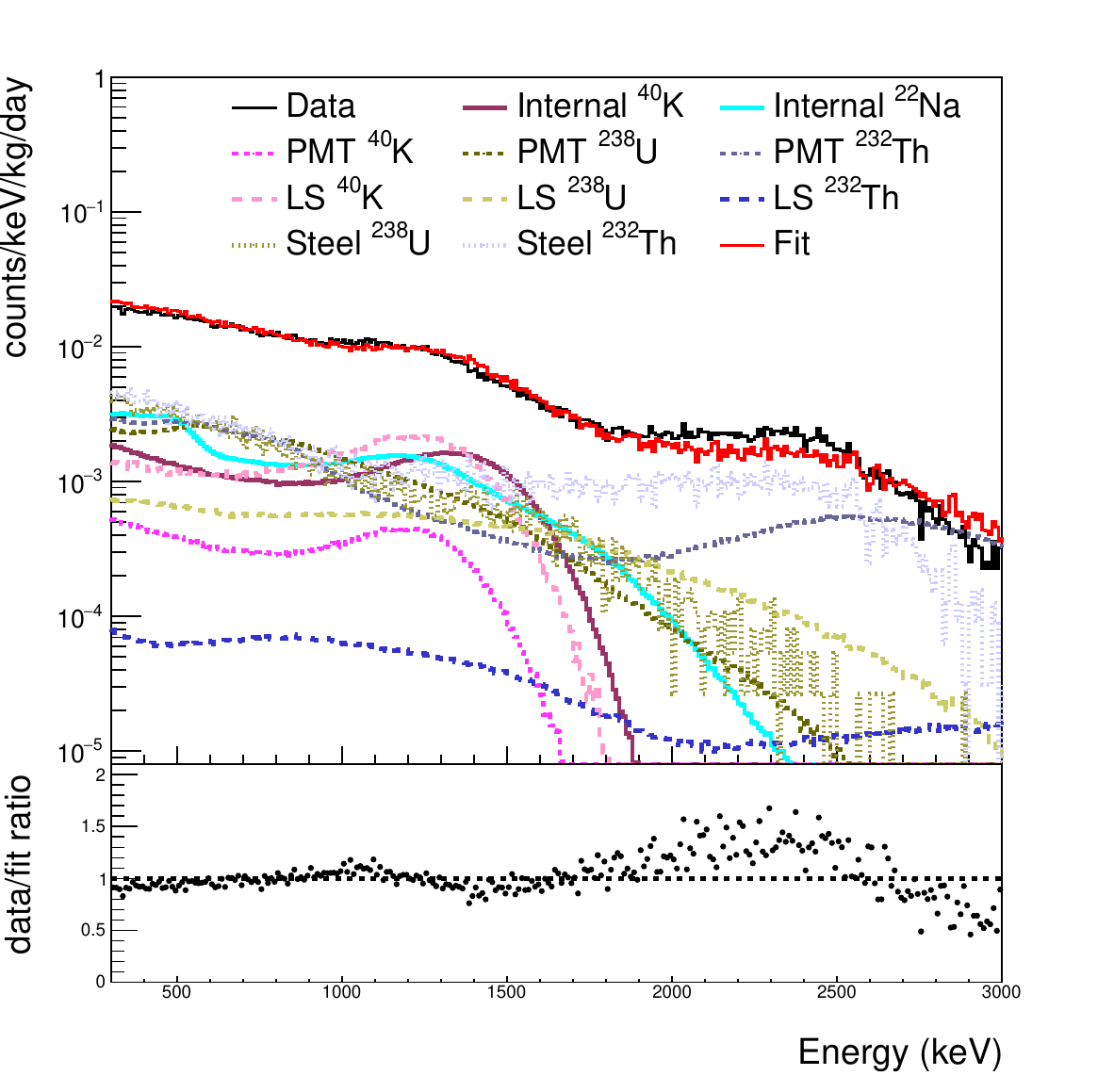}
	\caption{Fit of the first 100 hours of liquid scintillator data with Geant4 simulations modeling the main background components in the detector. Only the main internal and external components are considered in this fit. The bottom panel shows the ratio of the data to the total fit.}
	\label{fig:MCfit}
\end{figure}

The fit considers prior estimates of the activities of various background components, given in~\cite{COSINE_background}, as initial parameters in the fit. The $^{40}$K background in the LS is fit freely and later compared to recent measurements. Only 100 hours of data are used in the fit, to avoid spectral changes due to small gain drifts that occur over longer periods of time. This observed gain drift and its correction is explained in Sec.~\ref{sec:gain_drift}. The fitted spectrum can be seen in Figure~\ref{fig:MCfit}, while the best-fit background levels are given in Table~\ref{tab:backgroundFit}. The observed discrepancies between the data and MC indicates that both the resolution measurement and the modeling of external components needs to be improved. However, the primary goal of this study is not the measurement of the background activities, especially because of the challenges of performing an accurate calibration and mapping of the energy resolution of the LS. Nonetheless, the background modeling performed helped us better identify the most prominent components observed by the LS veto and gave an estimation of the $^{40}$K activity in the LS.

% \begin{table}[!htp]
% 	\centering
% 	\caption{Background fitting results.}
% 	\vspace{0.3cm}
% 	\label{tab:backgroundFit}
%     %\resizebox{width=0.8\textwidth}{!}{
% 	    \begin{tabular}{lc}
% 		    \hline
% 		    \hline
% 		    Component & Background (mBq/kg)  \\
% 		    \hline
% 		    Internal $^{40}$K & $<$ 17.9\\
% 		    Internal $^{22}$Na & $<$ 12.4\\
% 		    PMT $^{238}$U & $<$ 17.3\\
% 		    PMT $^{232}$Th & $<$ 130.3\\
% 		    PMT $^{40}$K & $<$ 51.6\\
% 		    LS $^{40}$K & $<$ 60.0\\
% 		    LS $^{238}$U & $<$ 0.17\\
% 		    LS $^{232}$Th & $<$ 0.04\\
% 		    External Steel $^{238}$U & $<$ 375\\
% 		    External Steel $^{232}$Th & $<$ 222\\
% 		    \hline
% 		    \hline
% 	    \end{tabular}
%     %}
% \end{table}	

\begin{table}[!h]
	\centering
	\caption{Liquid scintillator background fit results.}
	\vspace{0.3cm}
	\label{tab:backgroundFit}
  %\resizebox{width=0.8\textwidth}{!}{
	    \begin{tabular}{lc}
		    \hline
		    \hline
		    Component & Background (mBq/kg)  \\
		    \hline
		    Internal $^{40}$K & 7.04 $\pm$ 10.9\\
		    Internal $^{22}$Na & 0.63 $\pm$ 11.8\\
		    PMT $^{238}$U & 3.84 $\pm$ 13.5\\
		    PMT $^{232}$Th & 75.4 $\pm$ 54.9\\
		    PMT $^{40}$K & 32.5 $\pm$ 19.1\\
		    LS $^{40}$K & 5.9 $\pm$ 54.0\\
		    LS $^{238}$U & 0.09 $\pm$ 0.08\\
		    LS $^{232}$Th & 0.02 $\pm$ 0.02\\
		    External Steel $^{238}$U & 29.4 $\pm$ 346.0\\
		    External Steel $^{232}$Th & 49.1 $\pm$ 173.2\\
		    \hline
		    \hline
	    \end{tabular}
    %}
\end{table}

\section{Analysis Threshold Estimation}\label{sec:threshold}

Because of the relative rarity of WIMP interactions with ordinary matter, the application of a coincidence or anti-coincidence requirement between the sodium iodide detectors or the LS veto provides a key event discriminator between a possible WIMP signal and background events. Additionally, we can take advantage of these coincidence conditions to identify particular radioisotopes contaminating the detector. An event that activates a single NaI(Tl) detector is referred to as a single-hit event; events that activate multiple NaI(Tl) detectors, or that activate any number of NaI(Tl) detectors plus the LS veto, are classed as multiple-hit events. Events identified as muons are typically vetoed and not subject to the single- or multi-hit classification. An event occurring in the LS veto is defined as activating the veto if the measured energy deposited in the LS veto is above a set threshold. We seek an energy threshold at which the rate of LS veto events results in a deadtime of less than 0.5\%, which is to minimize noise event contamination in the event selection.

\begin{figure}[!h]
	\centering
	\includegraphics[width=0.5\columnwidth]{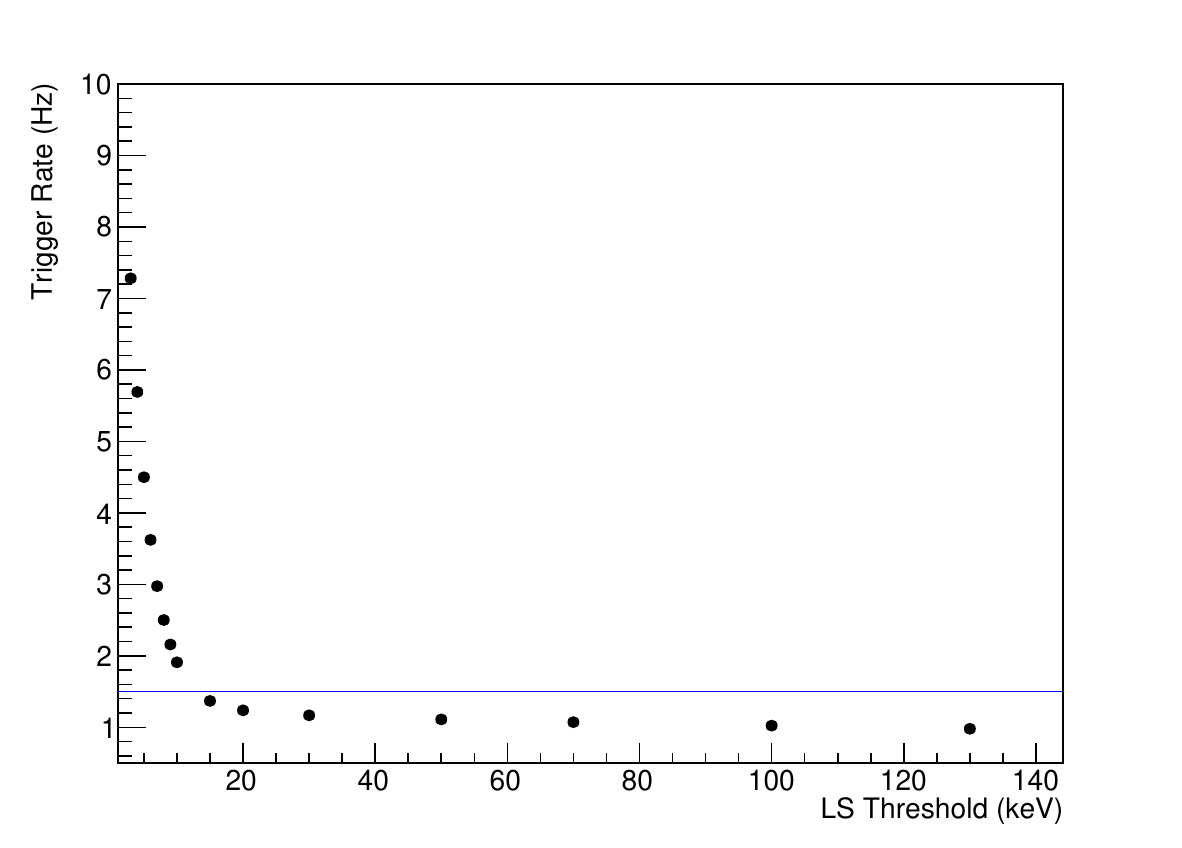}
	\caption{LS trigger rate as a function of LS energy threshold. A horizontal line is drawn at 1.5~Hz, which equates to a 0.5\% deadtime of the LS DAQ.} 
	\label{fig:deadtime}
\end{figure}

In order to determine the LS veto energy threshold, we perform a conservative calculation of the LS veto-induced deadtime. The conservative nature of this approach originates with a worst-case assumption: we assume LS veto events are non-overlapping; that is, we assume that no LS veto events occur within the deadtime of another LS veto event. This minimizes the number of events needed to achieve a specific deadtime. Under this assumption, and the fact that the duration of a single LS veto event is 200~ns, an event rate of 100 Hz would lead to a 20~$\mu$s/second, or 0.002\%, deadtime under these conservative assumptions ~\cite{COSINE_daq}. In addition, we must take into account the fact that for non-muon events the LS veto is passively triggered and is thus only acquires data when activated by a trigger from one of the NaI(Tl) detectors. Following a NaI(Tl) trigger, the LS veto acquires data for a 4~$\mu$s period surrounding this trigger. Thus, our observed 15~Hz average NaI(Tl) trigger rate results in a 60~$\mu$s/second livetime for the LS veto. A 0.5\% deadtime within this period then corresponds to 300 ns per 60~$\mu$s. As each LS veto event occupies 200~ns, we require a LS veto event rate equal to or below 1.5~Hz in our worst-case assumption. Figure~\ref{fig:deadtime} shows how the LS total trigger rate changes as a function of the energy threshold. With a higher energy threshold, the LS trigger rate decreases as fewer events pass that energy threshold. This demonstrates that by having a conservative threshold of 20~keV, you can achieve $<$0.5\% deadtime in the LS channels.

\section{Background Tagging Efficiency}\label{sec:efficiency}
\begin{figure}[!h] 
    \centering
    \includegraphics[width=0.9\columnwidth]{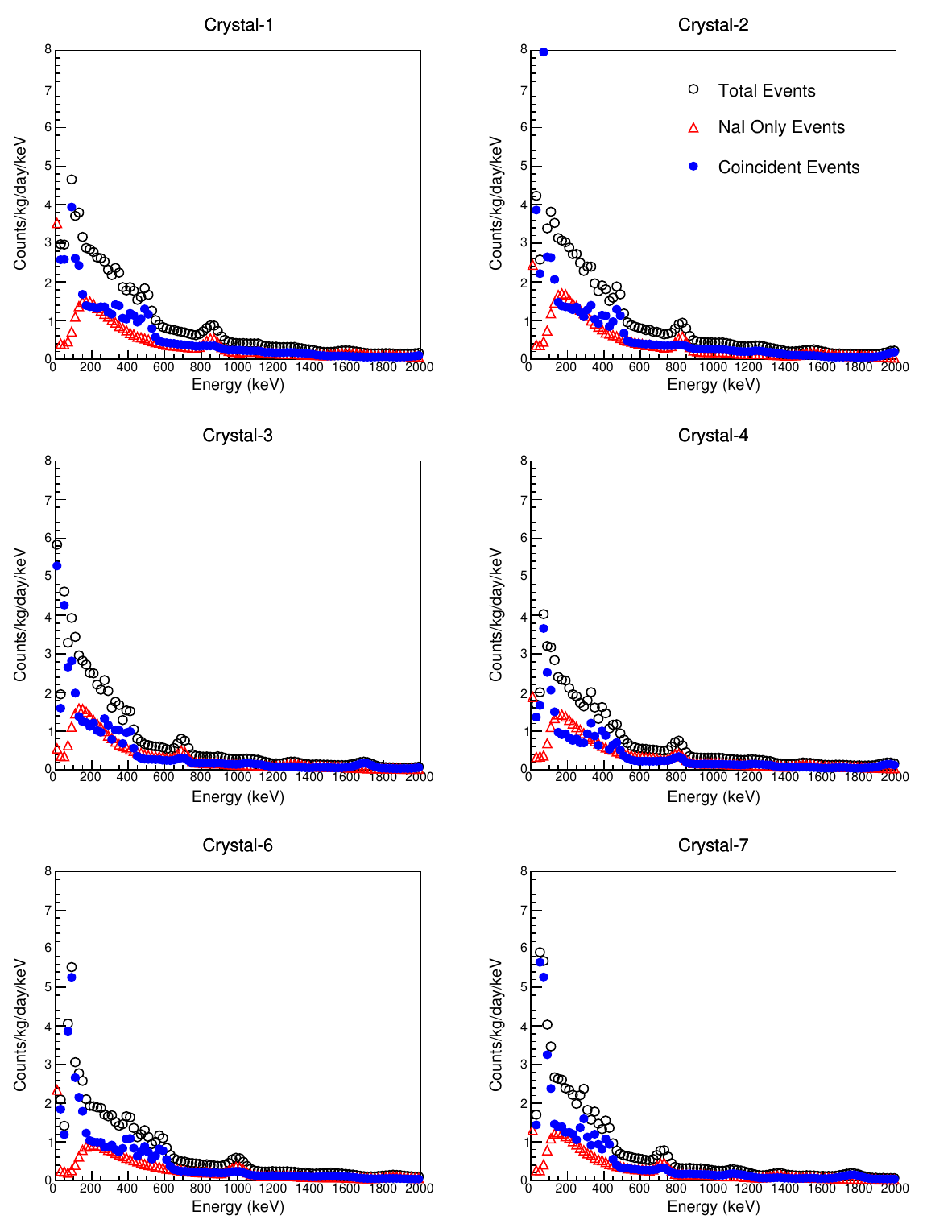}
    \caption{Background energy spectra of the six NaI(Tl) crystals in the COSINE-100 detector system at high energy. Vetoed events (blue filled circles) deposit energies greater than 20~keV in the LS veto, whereas events defined as NaI(Tl) only (red triangles) have no such depositions in the LS veto detector. The black, open circle spectrum represents all events (the sum of events in both categories). Crystal-5 and Crystal-8 are not shown due to their lower light yields and, hence, higher energy thresholds.}    
    \label{fig:spectrum_HE}
\end{figure}

\begin{figure}[!h] 
    \centering
    \includegraphics[width=0.9\columnwidth]{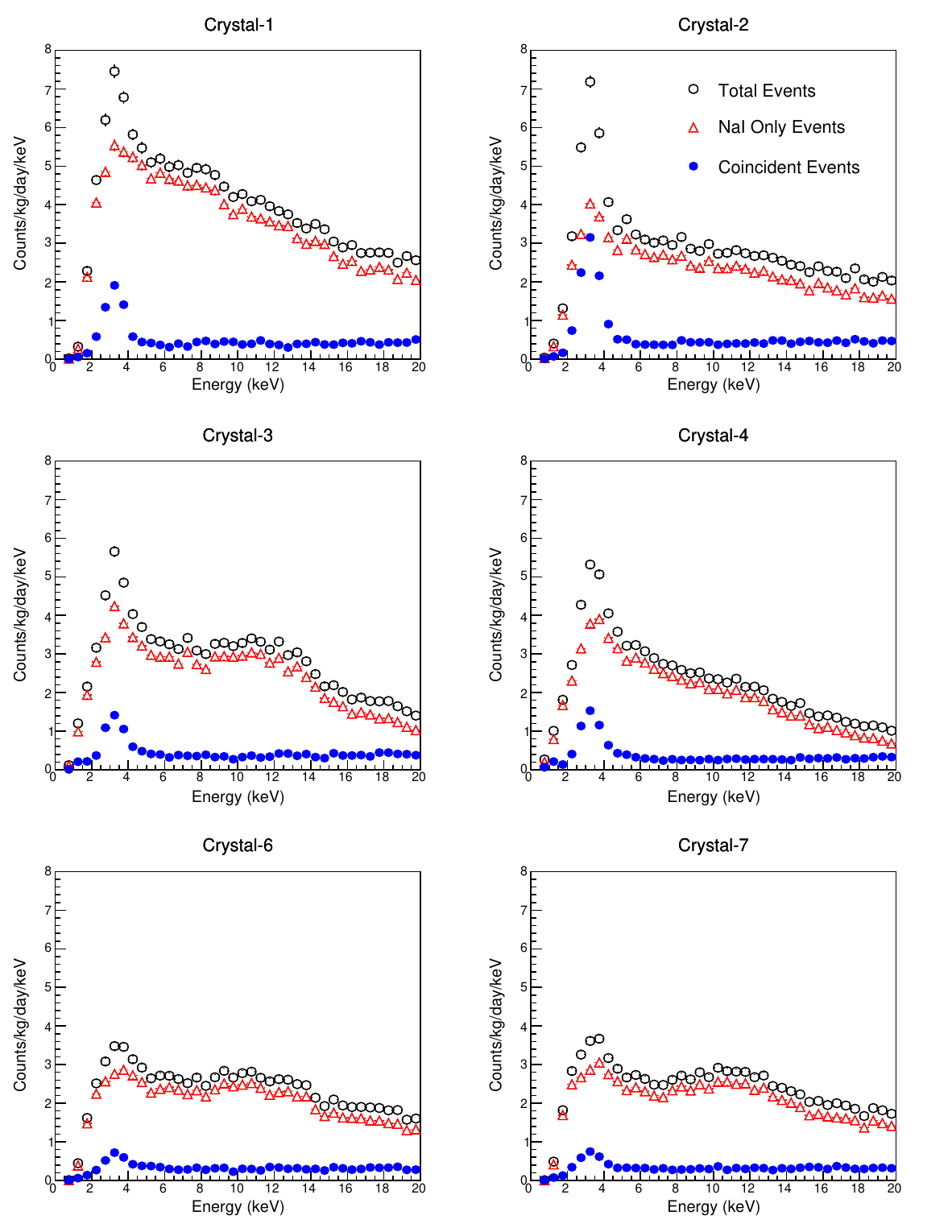}
    \caption{Background energy spectra of the six NaI(Tl) crystals in the COSINE-100 detector system at low energy. Vetoed events (blue filled circles) deposit energies greater than 20~keV in the LS veto, whereas events defined as NaI(Tl) only (red triangles) have no such depositions in the LS veto detector. The black, open circle spectrum represents all events (the sum of events in both categories). Crystal-5 and Crystal-8 are not shown due to their lower light yields and, hence, higher energy thresholds.} 
    \label{fig:spectrum_LE}
\end{figure}

The tagging efficiency is defined as the fraction of events tagged (vetoed) by the LS among all the triggered events in the crystals. In order to estimate the tagging efficiency of the LS system in COSINE-100, we measure the event rates of the NaI(Tl) detectors with and without the LS veto requirement, as shown in Figure~\ref{fig:spectrum_HE} and Figure~\ref{fig:spectrum_LE}. A significant reduction in the background rate of the NaI(Tl) detectors is realized by requiring that no signal be present in the LS veto detector. The total tagging efficiency of events coincident with the six active crystals in the COSINE-100 system, which is calculated using the total number of LS-tagged events, is shown in Table~\ref{tab:taggingEfficiencyTotal}, for 2--6, 6--20, and 100--1500~keV energy regions as representative regions. Two crystals, Crystal-5 and Crystal-8, are not shown due to their lower light yields and, hence, higher energy thresholds (4~keV and 8~keV)~\cite{COSINE_detector}. The total tagging efficiencies of the LS are between 10\% and 16\% in the low energy region and between 50\% and 62\% in the high energy region, for different crystals in the detector system. This is clearly observed in Figure~\ref{fig:spectrum_HE} and Figure~\ref{fig:spectrum_LE} by comparing the rates of coincident events and total events. The higher tagging efficiency at high energies is because there is a relatively higher contribution from radioactive isotopes at higher energies, such as nuclei in the $^{238}$U and $^{232}$Th decay chain. The difference in the tagging efficiency between different crystals is mainly due to each crystal's different background compositions. The tagging efficiencies in the energy interval of 2--6~keV are higher than those in the 6--20~keV interval, which indicates that the tagging efficiency of $^{40}$K is higher than that of most other components.

\begin{table*}[!b]
	\centering
	\caption{Total tagging efficiencies for six crystals in the COSINE-100 detector system calculated from data.}
	\vspace{0.3cm}
	\label{tab:taggingEfficiencyTotal}
    \resizebox{\textwidth}{!}{
	    \begin{tabular}{ccccccc}
		    \hline
		    \hline
		    Total tagging efficiency (\%) & Crystal-1 & Crystal-2 & Crystal-3 & Crystal-4 & Crystal-6 & Crystal-7 \\
		    \hline
		    2--6~keV      &  14.75 & 28.90 & 17.34 & 18.48 & 14.66 & 14.30 \\
		    6--20~keV     &  10.88 & 16.52 & 13.98 & 14.39 & 13.06 & 13.15 \\
		    100--1500~keV &  57.53 & 54.69 & 50.92 & 49.79 & 61.92 & 57.68 \\
		    \hline
		    \hline
	    \end{tabular}
    }
\end{table*}	

\begin{table*}[!b]
    \centering
	\caption{$^{40}$K tagging efficiencies for six crystals estimated using a Geant4 simulation.}
	\vspace{0.3cm}
	\label{tab:taggingEfficiencyK40}
	\resizebox{\textwidth}{!}{
	    \begin{tabular}{ccccccc}
		    \hline
		    \hline
		    $^{40}$K Tagging efficiency (\%) & Crystal-1 & Crystal-2 & Crystal-3 & Crystal-4 & Crystal-6 & Crystal-7 \\
		    \hline
		    2--4~keV &  75.12 & 72.72 & 69.08 & 75.94 & 66.52 & 65.80 \\
		    2--6~keV &  73.51 & 71.16 & 67.54 & 74.46 & 65.18 & 64.39 \\
		    0--10~keV & 71.35 & 69.02 & 65.56 & 71.58 & 63.08 & 62.42 \\
		    \hline
		    \hline
	    \end{tabular}
	}
\end{table*}

One of the most prominent backgrounds for dark matter searches with NaI(Tl) crystals is the internal $^{40}$K decay to $^{40}$Ar~\cite{COSINE_detector,BarbosaDeSouza:2017,COSINE_WIMP,ANAIS2014,ANAIS2016,SABRE}. This decay generates an X-ray at approximately 3.2~keV with a 1,460~keV $\gamma$-ray. If the accompanying 1,460~keV $\gamma$-ray escapes from the crystal, the event consists of a single 3.2~keV deposition, which becomes a background in the region of interest for the WIMP search. However, if we tag the escaping 1,460~keV $\gamma$-ray with the LS veto detector, we can tag the 3.2~keV X-ray in the NaI(Tl) crystal as a non-WIMP event. 

In order to properly obtain the $^{40}$K tagging efficiency of the LS system, we need to simulate various background components with the Geant4 simulation package. The veto efficiency for energies between 2--4~keV is estimated from the internal $^{40}$K simulation, which shows that the tagging efficiency of the $^{40}$K events is 65 -- 75\%. Table~\ref{tab:taggingEfficiencyK40} summarizes the $^{40}$K tagging efficiencies for the six crystals considered in this analysis. The untagged $^{40}$K events are due to the 1,460~keV $\gamma$-rays escaping the LS veto without scatter, scattering off another volume within the LS veto, such as the crystal encapsulation, or due to $^{40}$K decaying into $^{40}$Ca instead of $^{40}$Ar. The tagging efficiencies for $^{40}$K decays originating in Crystal-1 and Crystal-4 are found to be higher than the other crystals since these two crystals are located at the edges of the array, and thus surrounded by fewer non-scintillating detector components. The efficiencies of Crystal-6 and Crystal-7, on the other hand, are the lowest as they are near the bottom of the LS detector and are also surrounded by neighboring crystals. The filled blue circles in Figure~\ref{fig:spectrum_LE} show the low-energy spectrum of the NaI(Tl) crystal that is tagged by the LS veto detector. The size of the 3.2~keV peak from $^{40}$K decay differs based on the crystal due to differences in initial $^{40}$K contamination levels in each of them, with Crystal-2 having the highest level of $^{40}$K contamination, followed by Crystal-1/Crystal-3/Crystal-4 and Crystal-6/Crystal-7~\cite{COSINE_detector}.

\section{Gain fluctuation and correction}\label{sec:gain_drift}
\begin{figure}[t]
\centering
    \subfloat[LS gain vs. time before correction]{
        \centering
        \includegraphics[width=0.5\columnwidth]{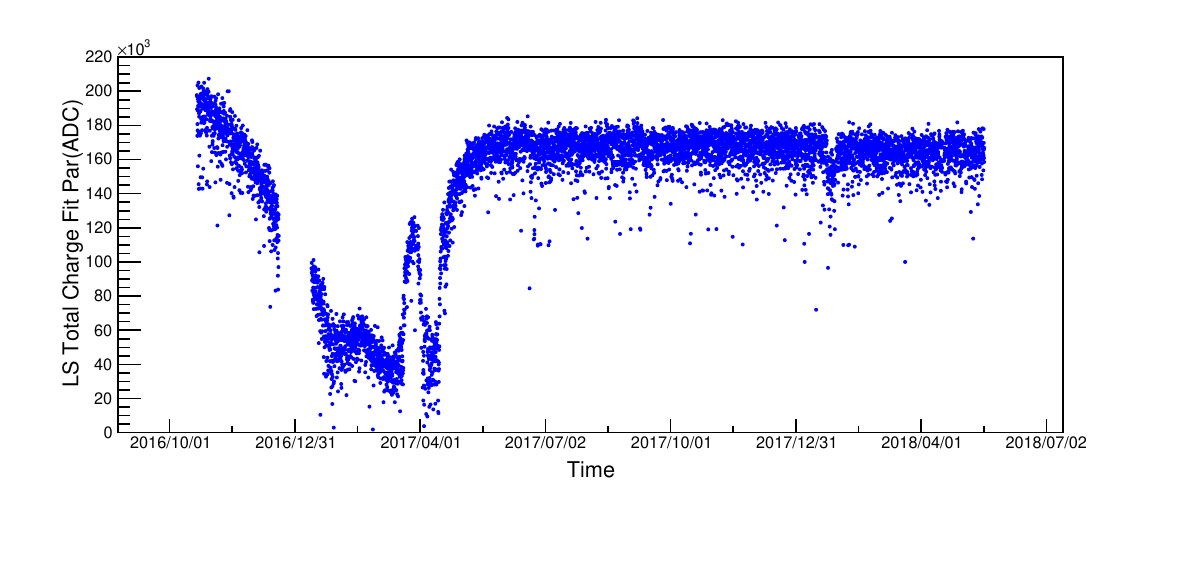}
        \label{fig:lsgain_before}
    }
    \\
    \subfloat[LS gain vs. time after correction]{
        \centering
        \includegraphics[width=0.5\columnwidth]{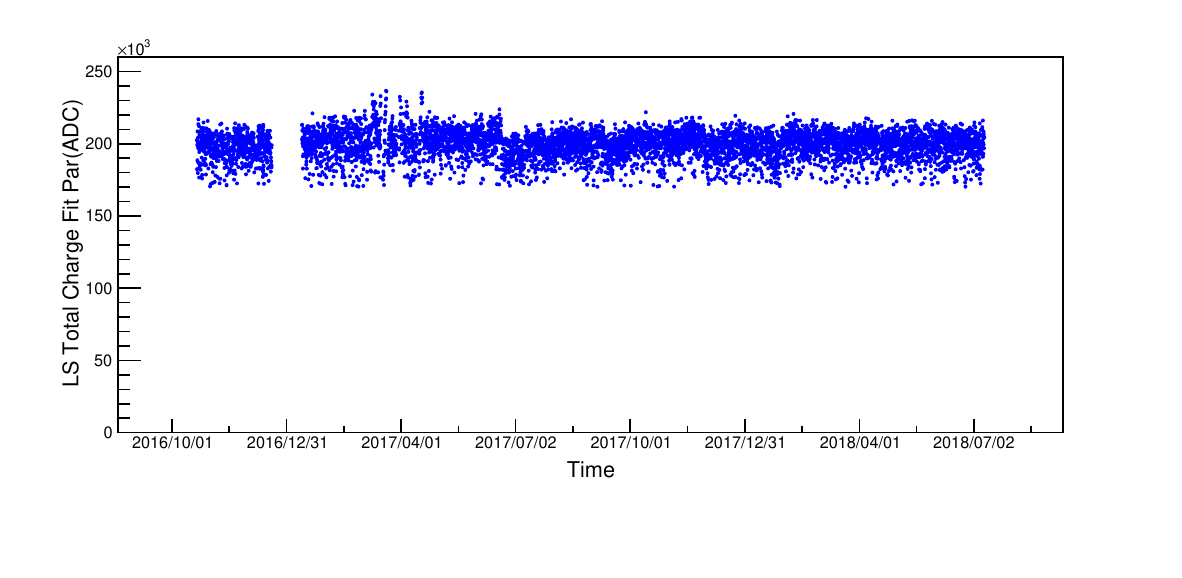}
        \label{fig:lsgain_after}
    }
\caption{LS gain fluctuation as a function of time (a) before and (b) after correction. The gain of the LS is represented by the location of the Compton edge of 1,460 keV $\gamma$-ray scatters, which originate from $^{40}$K decays.}
\end{figure}

Fluctuations in the LS veto gain can be monitored by tracking the Compton edge of $^{40}$K over time. To quantitatively evaluate the LS veto gain fluctuations, we fit the Compton edge of the 1,460 keV $\gamma$-ray found in the LS spectrum with the following phenomenological function,
\begin{equation}\label{eq:fit}
    y = \frac{p_{0}}{e^{p_{1}(x-p_{2})}+1},
\end{equation}
and extract the $p_{2}$ parameter, which represents the location of the Compton edge. The plot of the parameter $p_{2}$ as a function of time is shown in Figure~\ref{fig:lsgain_before}. We was observed that the gain slowly decreased from the beginning of the physics run, hitting a minimum in early April 2017. The gain then increased to ~85\% of its original value around late June 2017, and has been slowly decreasing since.

We have not yet found the cause of this behavior, as the current DAQ configuration, which only saves the charge sum of a triggered event, only provides limited information about the behaviour of the LS veto system. Nonetheless, it is important to correct for gain shifts in the LS veto as it will impact the signal event selection of COSINE-100 if untreated. We used the fitted $p_{2}$ parameter from the LS spectrum of the first dataset of the physics run as a reference value and corrected each subsequent dataset by applying correction factor to match the the $p_2$ parameter. The $p_2$ parameter from the corrected LS spectrum as a function of time for each dataset can be found in Figure~\ref{fig:lsgain_after}, which illustrates the increase LS veto energy scale stability provided by the gain correction. 

\section{Conclusion}\label{sec:conclusion}
The purpose of the liquid scintillator veto system of the COSINE-100 experiment is to reduce the background level both from the external and internal sources. The liquid scintillator is carefully produced to have a high light yield with a low radioactive background, and a dataset with a total exposure of 59.5 live days was used for this analysis. A Geant4 simulation shows a reasonable agreement with the data measured from the LS system, and this simulation result is then used to cross-check the calibration of the LS. The analysis threshold is determined to minimize noise event contamination in event selection. The tagging efficiency of 3.2~keV X-rays from internal $^{40}$K decays is estimated to be about 65 to 75\% from the Geant4 simulation. 

\section*{Acknowledgements}
We thank the Korea Hydro and Nuclear Power (KHNP) Company for providing underground laboratory space at Yangyang.
This work is supported by:  the Institute for Basic Science (IBS) under project code IBS-R016-A1 and NRF-2016R1A2B3008343, Republic of Korea;
NSF Grants No. PHY-1913742, DGE-1122492, WIPAC, the Wisconsin Alumni Research Foundation, United States; 
STFC Grant ST/N000277/1 and ST/K001337/1, United Kingdom;
Grant No. 2017/02952-0 FAPESP, CAPES Finance Code 001, CNPq 131152/2020-3, Brazil.

\bibliographystyle{unsrt}
\bibliography{ref}

\end{document}